\documentstyle[12pt]{article}

\makeatletter\@addtoreset{equation}{section}\makeatother

\newcommand{\be}{\begin{equation}}
\newcommand{\ee}{\end{equation}}
\newcommand{\bea}{\begin{eqnarray}}
\newcommand{\eea}{\end{eqnarray}}

\newfont{\Bbb}{msbm10 scaled 1200}     
\newcommand{\mathbb}[1]{\mbox{\Bbb #1}}

\input epsf

\newcommand{\sectiono}[1]{\section{#1}\setcounter{equation}{0}}

\newcommand{\refb}[1]{(\ref{#1})}

\renewcommand{\title}[1]{\vbox{\center\LARGE{#1}}\vspace{5mm}}
\renewcommand{\author}[1]{\vbox{\center#1}\vspace{5mm}}

\newcommand{\email}[1]{\vbox{\center\tt#1}\vspace{5mm}}

\def\sqr#1#2{{\vcenter{\vbox{\hrule height.#2pt
         \hbox{\vrule width.#2pt height#1pt \kern#1pt
            \vrule width.#2pt}
         \hrule height.#2pt}}}}

\begin{document}

\begin{titlepage}

{}~ \hfill\vbox{\hbox{hep-th/0403153} \hbox{PUPT-2117} }\break

\vskip 1cm

\begin{center}
\large{\bf  The Horizon Order Parameter }

\vspace{10mm}




\normalsize{N. Itzhaki}

\vspace{10mm}



\normalsize{\em Joseph Henry Laboratories, Princeton University,\\
Princeton, NJ 08544, USA}


 \email{nitzhaki@princeton.edu}

\vspace{5mm}

\end{center}

\begin{abstract}


We construct a diffeormorphism invariant  operator that is
sensitive to how far we are from the black hole horizon. Its
expectation value  blows up on the horizon and it is small away
from the horizon. Using this operator, we propose a non-standard
effective action that, we argue, captures some of the relevant
physics of quantum black holes, including the absence of the
horizon at the full quantum level. With the help of a toy version
of this effective action, we speculate on a possible connection
between UV/IR mixing and the cosmological constant problem.

\end{abstract}

\end{titlepage}

\baselineskip=18pt

\newpage

\sectiono{Introduction}

Recent progress in string theory suggests that information is
preserved in the process of black hole formation and evaporation.
The reason is that   in some cases there is a dual description for
this process in terms of gauge theories that are believed to be
unitary. This progress, however, does not clarify the  mechanism
that extracts the information back to infinity. In particular, we
still do not known  what is wrong with Hawking's original argument
\cite{ha} that the information is lost. Does the horizon or the
singularity (or perhaps  both) play the crucial role in the
resolution of the black hole information puzzle? It cannot be the
singularity since  by the time we approach the singularity the
information is already lost.\footnote{A possible way to overcome
this was proposed recently in \cite{hm}.} It cannot be the horizon
since locally there is nothing special about the horizon as all
curvature invariants on the horizon are small. This is  the
reasoning that led Hawking to conclude that the information is
lost.

The possibility that, despite the small  curvature, something
dramatic does happen on the horizon was studied by several authors
(see e.g \cite{th}-\cite{ma2}). The argument  is as  follows. For
a large black hole the energy of a typical Hawking particle is
small at infinity - it is of the order of $1/M$. In normal
situations this  means that its backreaction is negligible.
However, in the black hole case the energy is blue shifted by an
infinite amount when traced back to the horizon. Thus no matter
how large the black hole is this energy is blowing up  close
enough to the horizon. This is known as the transplanckian
problem. The simplest and possibly strongest argument against this
is that if the horizon is indeed a special hypersurface, then we
should be able to write down an effective action that differs
substantially from GR on the horizon and is well approximated by
GR away from the horizon. But again all curvature invariants are
small on the horizon so such an effective action cannot exist.

The standard answer to this criticism is that the effective action
we are after must be non-local, and hence  hard to find. From the
gravity point of view the  non-locality follows from the
definition of the horizon. Roughly speaking, the horizon is
defined as the boundary between the region from which massless
particles can reach infinity and the region from which they
cannot. This is  not a local definition. From  a field theory
perspective this effective action, in principle, can be derived by
integrating out the radiation. Since most of the radiation is of
massless particles  the effective action  is non-local, and  must
involve more than just local curvature invariants.

Without a  concrete non-local action, it is hard to test this
point of view or to see whether it can be useful for other issues
in quantum gravity (e.g., the cosmological constant problem). The
goal of this paper is to try to improve on that answer by offering
an effective action that captures at least some of the relevant
physics. The key ingredient that we introduce in section 2 is the
horizon order parameter. The horizon order parameter is a
diffeormorphism invariant operator whose expectation value is
sensitive to how far we are from the horizon. In particular, it
blows up on the horizon. We show that with spherical symmetry such
an operator can be defined locally. In the general case one cannot
find a local horizon order parameter. Instead what we  do in this
case is to trade the non-locality with a new local field that we
``integrate in". In terms of this field and the metric, the
horizon order parameter we propose is local.

Equipped with the horizon order parameter, we consider in section
3 a couple of possible actions that have the desirable properties
described above. In section 4 we compare these scenarios with
experiments. In section 5 we discuss a possible relation with the
cosmological constant problem.

\sectiono{The horizon order parameter}

\subsection{The spherical symmetric case}

As stressed in the introduction the horizon order parameter cannot
be  a local operator of the metric. With spherical symmetry,
however, a local operator contains more information than it
normally does. By measuring the expectation value of some gauge
invariant operator at $r, \theta, \phi$ we know its value
 at the same $r$
but for any $\theta$ and $\phi$. This trivial statement combined
with  Gauss's law suggests that for spherically symmetric
configurations the horizon order parameter can be a local
operator.

Let us first recall the usual argument why such an operator should
not exist. Consider a four dimensional Schwarzschild black hole
with the familiar metric
\begin{equation}\label{01}
ds^2=-(1-\frac{2M}{r})dt^2+\frac{dr^2}{1-\frac{2M}{r}}+r^2
d\Omega_2^2,
\end{equation}
and define a new radial coordinate that measures the invariant
distance from the horizon
\begin{equation}\label{0}
\rho =\int_{2M}^r dr \sqrt{g_{rr}}\approx\sqrt{8M(r-2M)}.
\end{equation}
For $\rho\ll M$ we find that \refb{01} can be approximated by
Rindler space
\begin{equation}\label{02}
    ds^2=-\frac{\rho^2}{16M^2} dt^2 +d\rho^2 +dY^2 +dZ^2.
\end{equation}
This means that  there can be  no diffeormorphism invariant
operator that depends on $\rho$ because Rindler space is Minkowski
space written in an accelerating coordinates.  As we increase $M$
the region Rindler space approximates well, is getting larger.
Thus it seems  that at least for a large black hole there can be
no horizon order parameter.

Rindler space, however, is only an approximation to the near
horizon geometry. For example, the expectation value of the
following operator
\begin{equation}\label{506}
    N=R_{\mu\nu\rho\sigma}R^{\mu\nu\rho\sigma},
\end{equation}
vanishes  in Minkowski, while in the Schwarzschild background it
is
\begin{equation}\label{03}
     \langle N\rangle  =48\frac{M^2}{r^6}.
\end{equation}
For a large black hole $N$ is small on the horizon. Moreover, $N$
 is a smooth function of $r$
with no special features at $r=2M$. Therefore by measuring $N$ we
cannot tell whether we are approaching  the horizon of some black
hole. The same goes for any polynomial of the curvature.

Derivatives of the curvature are more interesting. Consider for
example the simplest operator that does not vanish in a Ricci flat
space
\begin{equation}\label{0}
\tilde{N}=
\nabla_{\beta}R_{\mu\nu\rho\sigma}\nabla^{\beta}R^{\mu\nu\rho\sigma}.
\end{equation}
The expectation value of that operator in the Schwarzschild
geometry is
\begin{equation}\label{4}
\langle \tilde{N} \rangle = 720\frac{ M^2(r-2M)}{r^9}.
\end{equation}
Much like \refb{03}, it is a smooth function of $r$. But, unlike
\refb{03}, here  something special happens on the horizon: outside
the horizon \refb{4} is positive, while inside the horizon it is
negative. On the horizon itself $\tilde{N}$ vanishes. This is not
an accident.  In static spherical symmetric situations the
curvature depends only on $r$, and $g^{rr}$ switches sign on the
horizon.

This suggests that $\tilde{N}$ can be used to construct the
horizon order parameter. Consider for instance
\begin{equation}\label{420}
{\cal O}_1=\frac{N^2}{\tilde{N}}.
\end{equation}
The expectation value of ${\cal O}_1$ in the black hole background
is
\begin{equation}\label{767}
   \langle {\cal O}_1\rangle =\frac{16 M^2}{5(r-2M)r^3}\approx \frac{1}{\rho^2}.
\end{equation}
Thus at any classical distance away from the horizon of a large
black hole, or  outside a massive star, it is much smaller than
one. But at Planckian distances from the horizon ($\rho\sim 1$) it
is of order one. On the horizon itself it blows up.

In Ricci curved spaces ${\cal O}_1$ has to be modified.
 For example in the case of
near extremal black holes ${\cal O}_1$ is of order $1$ at a
distance from the horizon that depends on the energy above
extremality.  It also blows up everywhere in AdS (and dS) spaces.
To fix this we define a slightly more complicated operator
\begin{equation}\label{o2}
{\cal O}_2=\frac{N C^2}{\nabla_{\beta}
C_{\mu\nu\rho\sigma}\nabla^{\beta}R^{\mu\nu\rho\sigma}},
~~~~C^2=C_{\mu\nu\rho\sigma}C^{\mu\nu\rho\sigma},
\end{equation}
where $C_{\mu\nu\rho\sigma}$ is the Weyl tensor. ${\cal O}_2$ also
works as it should in Ricci curved spaces and is equivalent to
${\cal O}_1$ in a Ricci flat space. It should be emphasized that
both ${\cal O}_1$ and ${\cal O}_2$ can be measured locally.

It is easy to verify that without spherical symmetry ${\cal O}_2$
does not work as it should. Namely, it blows up when it is not
supposed to and it does not blow up when it is supposed to. A
simple example is Kerr black hole. In that background ${\cal O}_2$
blows up on the ergosphere rather then on the horizon. A different
example that does not involve black holes  is of two massive
objects; along the line that connects their centers there is a
point where  ${\cal O}_2$ blows up.

\subsection{The general case}

Without spherical symmetry  the horizon order parameter cannot be
a local operator. It seems hopeless to guess a non-local operator
with the right properties. A more realistic approach
 is to add  new massless degrees of freedom. The horizon
 order parameter can then be local with respect to these new
degrees of freedom and the metric. In principle, one can integrate
out the new
 fields and express the order parameter as a  non-local operator
of the metric.

The challenge is  to find these new  massless degrees of freedom.
Recall   that there is a simple way to tell whether we are inside
or outside an eternal Schwarzschild black hole. There are four
Killing vectors: one in the Schwarzschild time direction and three
that are associated with rotations of
 the $S^2$. Inside the black hole all the Killing vectors  are
space-like while outside one of them is time-like. That Killing
vector is null on the Horizon. Therefore, the norms of the Killing
vectors contain the information on which side of the horizon we
are. In the case of a dynamical formation of black holes, in
general, there are no exact Killing vectors. But there  are
approximated Killing vectors.
 The no-hair theorem ensures that with the evolution of the
 collapsing matter they become increasingly  exact. Most importantly, the
approximated Killing vector that becomes null on the horizon
asymptotes to an exact Killing vector exponentially fast.
 This  plays a crucial role in  Hawking's derivation that the black hole
radiation depends only on the total mass, angular momentum and
charge, and not on the details of the matter that forms the black
hole.

This   suggests that the new degrees of freedom  we need to add is
a vector field $V^{\mu}$. Our task is to find an action for
$V^{\mu}$ whose equations of motions are solved by the vector
field described above. Namely, a vector field that asymptotes to a
Killing vector that becomes null on the horizon. We denote that
vector field by $\bar{V}^{\mu}$. The norm of $\bar{V}^{\mu}$ will
be used to construct the horizon order parameter. This  means that
$V^{\mu}$ is not  a gauge field. In order to find that action we
use the following properties of the Killing vectors in the black
hole geometry. First, $V^{\mu}$ is a Killing vector if the
following symmetric tensor vanishes
\begin{equation}\label{0}
G_{\mu\nu}=\nabla_{\mu}V_{\nu}+\nabla_{\nu}V_{\mu}.
\end{equation}
We do not want to impose $G_{\mu\nu}=0$ as a constraint since, as
we argue above,   in the generic case of black hole formation
there are no non-trivial solutions to that constraint. To find
approximated Killing vectors we need to find vectors that minimize
$G_{\mu\nu}$. The simplest action that does that is
\begin{equation}\label{s1}
S_1=- \int d^4x \sqrt{g} (G^2 + \lambda~ G_{\mu}^{\mu}).
\end{equation}
The constraint $G_{\mu}^{\mu}=0$ is needed to ensure that the
Hamiltonian in bounded from below.

 We still have to fix  $V^{\mu}$. A naive attempt to do this is to
impose that at infinity $V^2=-1$. That works  for the eternal
Schwarzschild black hole since  it fixes  \be\label{005}
V^{\mu}=(1,0,0,0),\ee in the coordinate system used in
eq.\refb{01}. In general, however, there are problems with that
proposal. The most obvious one is that   there is no reason why
the time-like Killing vector at infinity should agree with the
null Killing vector on the horizon. A simple  example is Kerr
black hole: At infinity the  Killing vector that satisfies
$V^2=-1$ is \refb{005}. But the Killing vector that becomes null
on the horizon of a Kerr black hole is
\begin{equation}\label{04}
V^{\mu}=(1,0,0,\Omega_H),
\end{equation}
where $\Omega_H$ is the horizon angular velocity that depends on
the mass and angular momentum of the black hole. That Killing
vector is actually space-like at infinity. Moreover, there are
situations in which imposing $V^2=-1$ as a boundary condition
simply does not make sense.

In short, the direction and magnitude of $V^{\mu}$  should be
determined by the dynamics at the neighborhood of the horizon and
not by the asymptotic region. One way to do this is to take
advantage of the fact that
 $\bar{V}^{\mu}$ is the only Killing vector for which
$\nabla^{\mu}(\bar{V}^2)$ is pointing, on the horizon, in the same
direction as $\bar{V}^{\mu}$ (see fig. 1). Then  one can define on
the horizon a scalar, $\kappa$, via the relation
\begin{equation}\label{0er}
\nabla^{\mu} (\bar{V}^2)=-2\kappa \bar{V}^{\mu},
\end{equation}
and show that $\kappa$ is a constant on the horizon (that is known
as the horizon surface gravity). In the normalization, that we do
{\em not} use, in which the $t$ component of $\bar{V}^{\mu}$ at
infinity is one, $\kappa$ is the temperature associated with the
black hole. The analog of the fact that $\kappa$ is constant on
the horizon is that the temperature is constant in thermal
equilibrium. This holds for any black hole, including black holes
with angular momentum.

Motivated by this we define the following vector field
\begin{equation}\label{540}
B^{\mu}= V^{\mu}+\nabla^{\mu} (V^2),
\end{equation}
and add to $S_1$
\begin{equation}\label{l}
S_2=\int d^4 x \sqrt{g} C^2\frac{B^4}{V^4}.
\end{equation}
There are several reasons for the appearance of  $C^2$ (that was
defined in \refb{o2}) in $S_2$.  First, we want to avoid
destabilization  of conformally flat spaces. Second, in the black
hole case $C^2\sim M^2/r^6$ and hence it suppresses the
contribution from the large $r$ region. Thus  $V^{\mu}$ is fixed
by the near  horizon region. Even in the neighborhood of  the
horizon $C^2\ll 1$. This  means that $S_2$ is a perturbation with
respect to $S_1$. This perturbation breaks the degeneracy among
the Killing vectors and fixes the direction and magnitude of
$V^{\mu}$.

To see this, note that if $V^{\mu}$ is space-like then $B^4>V^4$.
The reason is that for all Killing vectors outside the black hole
$\nabla^{\mu} (V^2)$ is a space-like vector that increases the
norm of $B^{\mu}$ relative to $V^{\mu}$. For the same reason, when
$V^{\mu}$ is time-like $B^4<V^4$. This fixes the direction of
$V^{\mu}$. The magnitude of $V^{\mu}$ is also fixed because the
first term in \refb{540} is linear in $V^{\mu}$ while the second
is quadratic in $V^{\mu}$. Indeed expanding \refb{l} around
$V^{\mu}=0$ in the black hole background, we find a tachyonic mode
that is stabilized when $B^2=0$. Near the horizon of a Kerr black
hole $B^2$ vanishes  for
\begin{equation}\label{t0}
V^{\mu}=\frac{1}{2\kappa}(1 ,0,0,\Omega_H),
~~~~~~\kappa=\frac{\sqrt{M^2-a^2}}{2M(M+\sqrt{M^2-a^2})}.
\end{equation}
It is important to emphasis that  while eq.\refb{0er} holds only
on the horizon itself , $B^2$ vanishes throughout the near horizon
region. Namely, $B^2\ll V^2$ as long as $\rho\ll M$ (see fig.1).

The direction of \refb{t0} is as expected. Since $\kappa$ is small
one might worry that the backreaction of $V^{\mu}$  is  too large
to be considered as a perturbation of the black hole geometry. The
$C^2$ in $S_2$ ensures that this does not happen and that the
energy in the $V^{\mu}$ field is much smaller than $M$.

\begin{figure}
\begin{picture}(240,232.5)(-40,-280)
\put(260,-100){$V^{\mu}$} \put(200,-143){$\nabla^{\mu}(V^2)$}
\put(50,30){\mbox{\epsfxsize=80mm \epsfbox{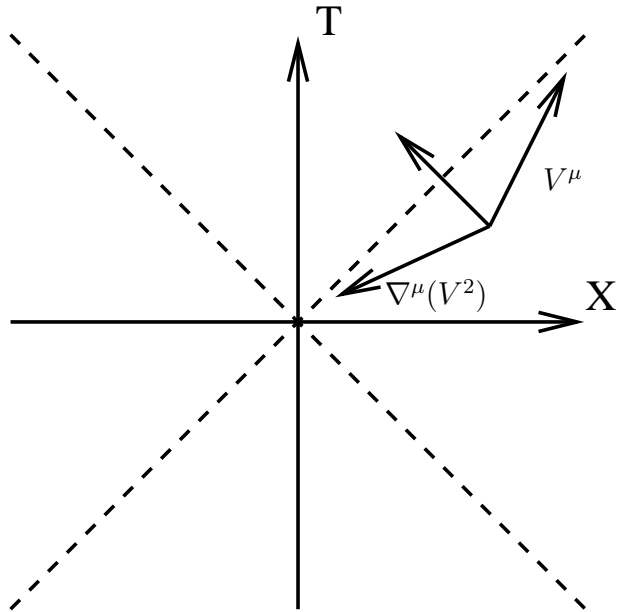}}}
\end{picture}
\caption{On the horizon (the dashed line) $V^{\mu}$ and
$\nabla^{\mu}(V^2)$ point in the same direction. Throughout the
near horizon region their sum is null. }
\end{figure}

The horizon order parameter is defined to be
\begin{equation}\label{or}
{\cal O}=\frac{F^2}{V^2},~~~~\mbox{where}~~~~
F_{\mu\nu}=\nabla_{\mu}V_{\nu}-\nabla_{\nu}V_{\mu}.
\end{equation}
On the horizon $V^2$ vanishes while $F^2$ is a constant (that is
related to the surface gravity) and therefore ${\cal O}$ behaves
as required. That is, ${\cal O}$ blows up as we approach the
horizon. For example in the  Schwarzschild case, $V^{\mu}$ is
given by \refb{t0} with $a=\Omega_H=0$ and   the expectation value
of ${\cal O}$ is
\begin{equation}\label{er}
\langle {\cal O}\rangle =\frac{2
M^2}{(r-2M)r^3}\approx\frac{1}{\rho^2}.
\end{equation}
Up to a constant of order one this expression is the same as
$\langle {\cal O}_1\rangle $ (see eq. \refb{767}). In particular,
it is much smaller than one at any classical distance from the
horizon, and at Planckian distances from the horizon it is of
order one. The advantage of ${\cal O}$ over ${\cal O}_1$ is that
for ${\cal O}$ this  also holds  in the absence of spherical
symmetry.

\sectiono{Effective actions}

Now that we have defined the horizon order parameter, we can
consider different effective actions that are sensitive to the
location of the black hole horizon. There are various interesting
scenarios  to study. Below we focus on the scenario that we
believe is the most plausible scenario.

\subsection{The no-horizon scenario}

Motivated by 't Hooft's S-matrix for black holes (for a review see
\cite{threv}) a scenario was proposed in \cite{it} (see also
\cite{en}) according to  which at the quantum level there is no
horizon to the black hole. The line of argument  that led to this
proposal is as follows. Using the conformal anomaly and
energy-momentum conservation one can calculate, subject to some
reasonable assumptions, the one-loop contribution to the
expectation value of the energy-momentum tensor associated with
the Hawking radiation \cite{un} (for a review see e.g. \cite{bi}).
The result of this calculation is in accord with Hawking's
assumption that for a large black hole nothing special happens on
the horizon. Namely, away from the horizon there is an out-going
flux of positive energy that is associated with the on-shell
Hawking radiation. Near the horizon, however, there is an in-going
flux of negative energy that corresponds to the partners of the
Hawking particles. As these particles are simply  falling through
the horizon their backreaction is regular. In fact, for massless
particles it is possible to  find the exact solution to Einstein's
equations near the horizon including the backreaction
\cite{massar}.

A classical system is fully described by  the one-point functions.
In a quantum theory, however, the one-point functions contain only
part of the information. Moreover, there are cases in which this
information is actually misleading. This was illustrated
beautifully in \cite{pol} where in a consistent quantum field
theory, states were constructed such that the one-point function
of the energy momentum tensor violates causality at arbitrarily
large distances.

It is certainly  possible that the  one-point functions of
\cite{un} are misleading as well.  This is supported by the
following argument. If the process of black hole creation and
evaporation is described by some complicated unitary S-matrix
amplitude then the out-going particles should be treated on equal
footing as the in-going particles. In particular, since the
ingoing particles are on-shell, the out going particles must be
on-shell as well. This implies that
 the out-going  radiation should be associated with a positive
out-going energy momentum flux everywhere, not just at infinity
(much like the flux  associated with the in-going particles is
on-shell everywhere). The backreaction in this case is such that a
test particle falling towards the black hole  will always be
causally connected with ${\cal I}^+$ \cite{it}. As strange as this
might seem, this is a consequence of  CPT; by time reversal this
statement is equivalent to the claim, that is obviously correct,
that the Hawking particles are causally connected with ${\cal
I}^{-}$. The result of \cite{un} is expected to be obtained by
first averaging over the final states and then calculating the
expectation value. Thus in this picture the black hole horizon is
an artifact of the semi-classical approximation that does not
exists at the full quantum level.

Concrete support for the no-horizon scenario comes from the D5-D1
system. This system is understood so well that a precise quantum
geometry can be associated with  each quantum state of the black
hole. Quite remarkably, Lunin and Mathur showed that these quantum
geometries do not admit horizons \cite{lm,lm2,ma3}. The horizon
appears only as a classical concept after averaging over the
quantum geometries. The techniques used in \cite{lm,lm2} are
special to this system and are valid when the black hole in
question is basically a BTZ black hole. Nevertheless, these
results clearly illustrate that the no-horizon scenario, as
radical as it might appear from a semi-classical point of view, is
a possibility that should be considered seriously.

As a first step towards realizing  the no-horizon scenario for
realistic black holes, it would be useful to have an effective
action that captures some of the relevant physics. With the help
of the horizon order parameter we can make some proposals. The
most straight forward one is
\begin{equation}\label{08p}
S=S_1+S_2+\int d^4 x \sqrt{g}\left( R +\exp({\cal O}) R^2+
\exp(2{\cal O}) R^3+...\right).
\end{equation}
This proposal is  inspired by string theory.   In string theory
the Planck scale is not a universal constant since it depends on
the local value of the dilaton. As a result the stringy loops
corrections to GR  depend on the dilaton. Here we have replaced
the dilaton with ${\cal O}$ because,  unlike the dilaton, the
expectation value of ${\cal O}$  blows up on the horizon. This
implies that the effective Planck scale grows as we approach the
horizon, and therefore the corrections to GR become increasingly
important. From eq.\refb{er} we see that at  a finite distance
from the horizon the effective Planck scale is of the order of the
size of the horizon itself. At that point all terms are equally
important. This  means that  the horizon becomes a non-local
sphere where the non-locally is of the order of the size of the
sphere itself. All the information is supposed to be encoded in
this non-local sphere.

Clearly there are many other actions with  similar properties. To
find the one that describes best the near horizon physics we have
to  make contact with the backreaction of the Hawking radiation
and with the black hole entropy. In our discussion we used
familiar ingredients, like surface gravity and null Killing
vectors, but we certainly have not made this connection precise.
It is likely that a better understanding of these issues will
modify, possibly substantially, the proposed horizon order
parameter and effective action.

An unusual feature of \refb{08p} is that it involves ratios of
polynomials of the fields. This is certainly beyond the Wilsonian
effective action framework, and it  raises many questions about
naturalness and quantum corrections. In particular, in our case,
unlike in the Wilsonian framework,  it is not even clear which
degrees of freedom are integrated out.
 This is closely related to the fact that at this stage, the precise
relation with the backreaction of the Hawking radiation and the
black hole entropy is unclear. At the moment we do not have good
answers to questions of this kind. The only comment we would like
to make is that the fact that we are forced to work outside the
framework of the Wilsonian effective action should not come as a
surprise. In the Wilsonian approach the number of degrees of
freedom is proportional to the volume of the system whereas
according to the holographic principle \cite{thhol,suhol} the
number of degrees of freedom should be proportional to the area
that surrounds the system.\footnote{The AdS/CFT correspondence
avoids this issue since the area of AdS is proportional to its
volume. In that regard the most concrete example we have so far to
the holographic principle is misleading.}  Put differently,  a
Wilsonian effective action  includes only polynomials of the
curvature and its derivatives. These are clearly not sensitive to
the location of the horizon. A closely related comment is about
UV/IR mixing. It is hard to imagine a realization of the
holographic principle without some kind of UV/IR mixing. Indeed
this is what is happening here. Naive power counting suggests that
the terms added to GR in \refb{08p} are irrelevant. Nevertheless,
as we approach the horizon they become as important as the
Einstein-Hilbert term. Of course this is not to say that the
scenario proposed here is the correct way to realize the
holographic principle, but it does seem to involve some of the
expected features.

An issue  that is often raised in relation to the no-horizon
scenario is that information can be  lost long before the horizon
is formed. Therefore the fact that an actual horizon is not formed
is not going to help with retrieving the information. Let us
illustrate this puzzle and its resolution in a Gedanken
experiment. Consider a spherically symmetric  null shell that
carries some information and is not energetic enough to form a
black hole by itself. In Fig.1 that shell is denoted by $A$. At a
later time there is an additional spherically symmetric null
shell, $B$, that has enough energy to form a black hole. By the
time the horizon is formed the information contained in $A$ is
already lost. Namely, point $c$ and ${\cal I}^+$ are not causally
connected.  In the no-horizon scenario shell $B$ follows the same
trajectory as before until it reaches $r=2M$. At this point the
shell becomes completely non-local, and it  floats while emitting
radiation which slowly decreases its mass. Shell $A$ reaches $r=0$
and as before it reexpands. However, instead of hitting a
space-like singularity it will hit shell $B$. At all times shell
$A$ is causally connected  with ${\cal I}^+$. A more general way
to say this, that is valid away from the thin shell limit, is the
following. Before reaching the singularity shell A has to cross
the apparent horizon. In the standard picture nothing special
happens there so the shell continues towards the singularity. In
the no-horizon scenario, however,   the apparent horizon and its
interior (including the would be singularity) are effectively
replaced by a giant  non-local ball.
\begin{figure}
\begin{picture}(240,232.5)(-40,-280)\put(100,-270){(a)}
\put(320,-270){(b)}\put(150,-55){${\cal I}^+$}
\put(310,-65){${\cal I}^+$}\put(115,-190){${\cal I}^-$}
\put(330,-210){${\cal I}^-$}
\put(40,20){\mbox{\epsfxsize=80mm \epsfbox{h.eps}}}
\put(250,30){\mbox{\epsfxsize=80mm \epsfbox{h1.eps}}}
\end{picture}
\caption{(a) The standard picture in which the information in the
shell A is lost long before the horizon was formed. (b) In the
no-horizon scenario shell A is always causally connected with
${\cal I}^+$. }
\end{figure}

\subsection{ Black hole as a superconductor}

 Equipped with
the horizon order parameter there are other types of effective
actions that one might want to explore. An especially entertaining
scenario is one in which the region behind the horizon is in a
different phase. In this scenario the black hole is the
gravitational analog of a superconductor: outside the black hole
we are in the Coulomb phase while the region behind the horizon is
in the Higgs phase. One way to realize the gravitational analog of
the Higgs phase is through  ghost condensation \cite{hamed}. We
want the kinetic term to have the wrong sign only inside the
horizon so  the action should have a term of the form
\begin{equation}\label{0i}
F({\cal O}) (\partial \phi)^2,
\end{equation}
where $F$ is some odd function of ${\cal O}$.  In this scenario
the information can fall behind the horizon, but since  the
dispersion relation in the  "Higgs" phase  is not  Lorentz
invariant \cite{hamed} the information can escape back to
infinity.

\sectiono{Comparison to experiment}

With the growing  evidence for black holes  one might suspect that
the no-horizon scenario  is ruled out by experiment. In this
section we briefly review these experiments and discuss their
relevance to our scenario.

The standard argument  for the existence of black holes is  that
there are dark objects that are simply too massive to be supported
by the nuclear force. Their mass is known by studying the
trajectories of nearby stars.  Since these trajectories are at
large distances from the horizon they do not involve large
gravitational field, and, in particular, they do not test the
no-horizon  scenario. There are, however, more recent observations
that do involve large red-shift and are more relevant to the
scenarios discussed here.

Some observations are based on the fact that the red-shift at the
location of the accretion disk for a Kerr black hole is
\begin{equation}\label{0}
1+z=\sqrt{\frac{1+a^2/r^2+2Ma^2/r^3}{1+a^2/r^2-2M/r}}.
\end{equation}
The Schwarzschild limit ($a=0$) is not very interesting in that
regard. The accretion  disk is located at  $r= 6M$ and we find
that the red-shift is pretty small, $z\sim 0.2$. More importantly,
this   probes  the geometry at $r\geq 6M$, and  not the
 near horizon geometry. In the extreme case ($a=M$) the
accretion disk is touching the horizon at $r=a=M$ and the maximal
red-shift is infinite. Experimentally  large red-shift in the iron
X-ray lines have been observed  in some Seyfert galaxies
\cite{iw}. For a large enough $a$ the accretion disk is located
between the ergosphere and the horizon. Therefore, these
observations could   rule out scenarios in which the order
parameter is ${\cal O}_2$ (or ${\cal O}_1$) and not ${\cal O}$.
This is in agreement with our theoretical expectations.

More evidence for black holes comes from the fact that type I
X-ray bursts are quite common in accreting neutron starts, but
have never been detected in accreting  black hole candidates (for
a recent review see \cite{bh}). This fits neatly with GR because
type I bursts are due to thermonuclear explosions taking place
when enough gas accretes on the surface of the   neutron star. For
black holes this does not  happen since there is no surface for
the gas to accrete on. In the scenario presented here the horizon
can be viewed as such a surface. Naively this suggests that
careful measurement of the X-ray burst can test the no-horizon
scenario. Unfortunately measurements of this kind cannot test this
scenario. The reason is simple. Indeed in this scenario the gas
accretes on the horizon. However, the local temperature on that
"surface" is of the order of the Planck scale that makes any
thermonuclear activity irrelevant.

\sectiono{The cosmological constant problem}

It is likely that a better understanding of black holes at the
quantum level will revolutionize our understanding of quantum
gravity in general and it might even shed new and unexpected light
on various issues in cosmology (for  suggestions see
\cite{fs,bf,bou}). In particular, an intriguing speculation is
that UV/IR mixing  plays an important role in the resolution of
the cosmological constant problem. The rough idea is that due to
large scale non-local effects, the macroscopic cosmological
constant might be much smaller than  the microscopic cosmological
constant. Where the macroscopic cosmological constant is the one
we observe, and the microscopic cosmological constant is
determined by the Planck or SUSY scale. Phrased slightly
differently that point of view has been advocated by Banks
\cite{banks1,banks2}.

Since the effective action we proposed involves UV/IR mixing it is
interesting to see if it provides  new insights to the
cosmological constant problem. Unfortunately, with our current
understanding of the horizon order parameter, it is premature to
have a rigorous   discussion on this issue. Instead, with the help
of a toy action, that is related to the actions considered above,
we illustrate, quite heuristically, how UV/IR mixing could have
interesting applications to  the cosmological constant problem.

The toy effective action we wish to consider is
\begin{equation}\label{opp}
\int d^4x \sqrt{g} (R+ R \frac{N^2}{\Box
N}+...),~~~~~N=R_{\mu\nu\rho\sigma}R^{\mu\nu\rho\sigma}.
\end{equation}
Since the second term is irrelevant with respect to the
Einstein-Hilbert term,  it is expected not to matter much when the
curvature is small in Planck units. Take for example a spatially
flat FRW universe
\begin{equation}\label{ppp}
    ds^2=-dt^2+a^2(t)\sum_{i=1}^3dx_i^2,
\end{equation}
with $a(t)\sim t^b.$ In that case  $R\sim 1/t^2$ whereas the
second term scales like $1/t^4$, and   as expected, it is
irrelevant at late times. In that respect, the second term in
\refb{opp} behaves  like an ordinary   higher loop correction to
GR. Furthermore, one can verify that the graviton propagator in
flat space is intact and that this effective action does not
contradict the solar system experiment.

The situation is quite different in a universe that is dominated
by a  cosmological constant. The solution to  Einstein's equations
in such a universe is a de-Sitter space-time. The usual higher
loops corrections are not going to change this. They could change
the exact relation between the cosmological constant and the
radius of curvature of de-Sitter. If the cosmological constant is
not Planckian but is fixed by the SUSY breaking scale, then even
the relation between the cosmological constant and the radius of
curvature is, to a good approximation, fixed by the Einstein's
equations. With the deformations considered here this is not the
case. Not only  we do not get de-Sitter as a solution, but the
solution we get is not highly curved. That is, the curvature of
space-time at late times is much smaller than $\Lambda$. The
reason is that although the second term in \refb{opp} is supposed
to be irrelevant,
 it actually blows up when the curvature is a non-zero constant.
 This
excludes de-Sitter and AdS \footnote{It is easy to generalize this
in a way that it vanishes for $AdS_5\times S^5$ when non-trivial
fluxes are involved. } as solutions regardless of how large the
cosmological constant is. Instead, we find a solution that
asymptotes at late times to
\be\label{097} a(t)=t^{{b(t)}},~~~~~b(t)=5/3-\frac{1}{\Lambda^2
t^{4}}. \ee
The  curvature  associated with that solution scales like
$1/t^{2}$. Thus at late times we get a smooth solution although
the cosmological constant is large. Put differently, \refb{097}
describes an accelerating universe, but the acceleration does not
depend on $\Lambda$ at all:
\begin{equation}\label{0i9}
\frac{\ddot{a}}{a}=\frac{10}{9t^2}.
\end{equation}
Notice that we did not have to introduce  a new scale  to the
action in order to get this.

Eq.\refb{0i9} describes a universe with $p/\rho =w= -0.6$. This
value is ruled out by the Supernova observation \cite{sn1,sn2}.
But, unlike the problems that will be  raised momentarily, this
should not be considered as a serious problem because other, more
complicated, actions yield an expansion that is consistent with
\cite{sn1,sn2}.

The  cosmological constant in the equation of motions is balanced
 not by a large curvature, as it is in GR, but rather by the fact that
the derivatives of the curvature invariants are even smaller than
the curvature itself. The  problem with that is that this
balancing is extremely fragile. There are other terms that can be
added to \refb{opp}, that are naively irrelevant, like $R
(\frac{N^2}{\Box N})^2$, but  change the solution \refb{097}
drastically. Thus we have not  made a real progress with the
naturalness of the cosmological constant. The only positive aspect
of this is that the corrected solutions  have at late times low
curvature as well.

There are other, more practical, problems  with this toy model.
Eq. \refb{097} is valid all the way to times as early as $\sim
1/\sqrt{\Lambda}$. This contradicts early universe observations.
For example,  the Hubble constant at nucleosynthesis is so
different than in the standard big-bang scenario that  it is hard
to see how abundances of light element could agree with
observation. This scenario is problematic even at late times. In
the presence of a large cosmological constant the equations of
motions are balanced by having a small $\Box N$. This has to be
done
 locally everywhere in the universe. But this is  inconsistent
with the solar system experiment, where $\Box N$ is determined by
the sun rather than by the cosmological evolution. Simply put,
 we indeed get a low curvature universe \refb{097} but there
is still a large energy density due to the cosmological constant.
This energy screens the gravitational effects of all other forms
of matter or radiations.

To conclude, this toy model has many problems and it is far from
solving  the cosmological constant problem. It does illustrate,
however, the role UV/IR mixing might play in  the resolution to
the cosmological constant problem.

\section{Summary}

The black hole information puzzle has generated over the years
fascinating ideas about quantum  black holes, large scale
non-locality  and holography. Without effective actions that
complement  these ideas it is hard to study or even state them
concretely. In this paper we attempted to make a small step
in that direction by constructing  an effective action that is
sensitive to the location of the black hole horizon.  We hope that
concrete progress will be made using actions of this kind.

\bigskip

\bigskip


\noindent {\bf Acknowledgements}

I  thank D. Kutasov for discussions. This material is based upon
work supported by the National Science Foundation under Grant No.
PHY 9802484. Any opinions, findings, and conclusions or
recommendations expressed in this material are those of the author
and do not necessarily reflect the views of the National Science
Foundation.


\begin{thebibliography}{99}

\bibitem{ha}
S.~W.~Hawking, ``Particle Creation By Black Holes,'' Commun.\
Math.\ Phys.\  {\bf 43}, 199 (1975).





\bibitem{hm}
G.~T.~Horowitz and J.~Maldacena, ``The black hole final state,''
arXiv:hep-th/0310281.





\bibitem{th}
G.~'t Hooft, ``On The Quantum Structure Of A Black Hole,'' Nucl.\
Phys.\ B {\bf 256}, 727 (1985).


\bibitem{thth}
G.~'t Hooft, ``The Black Hole Interpretation Of String Theory,''
Nucl.\ Phys.\ B {\bf 335}, 138 (1990).

\bibitem{Stephens:1993an}
C.~R.~Stephens, G.~'t Hooft and B.~F.~Whiting, ``Black hole
evaporation without information loss,'' Class.\ Quant.\ Grav.\
{\bf 11}, 621 (1994) [arXiv:gr-qc/9310006].




\bibitem{su}
L.~Susskind, L.~Thorlacius and J.~Uglum, ``The Stretched horizon
and black hole complementarity,'' Phys.\ Rev.\ D {\bf 48}, 3743
(1993) [arXiv:hep-th/9306069].



\bibitem{en}
F.~Englert, ``Operator weak values and black hole
complementarity,'' arXiv:gr-qc/9502039.

\bibitem{ver}
Y.~Kiem, H.~Verlinde and E.~Verlinde, ``Black hole horizons and
complementarity,'' Phys.\ Rev.\ D {\bf 52}, 7053 (1995)
[arXiv:hep-th/9502074].




\bibitem{it}
N.~Itzhaki, ``Is the Black Hole Complementarity principle really
necessary?,'' arXiv:hep-th/9607028.





\bibitem{ma22}
S.~D.~Mathur, ``Resolving the black hole information paradox,''
Int.\ J.\ Mod.\ Phys.\ A {\bf 15}, 4877 (2000)
[arXiv:gr-qc/0007011].

\bibitem{ma2}
S.~B.~Giddings and M.~Lippert, ``The information paradox and the
locality bound,'' arXiv:hep-th/0402073.


\bibitem{threv}
G.~'t Hooft, ``The scattering matrix approach for the quantum
black hole: An overview,'' Int.\ J.\ Mod.\ Phys.\ A {\bf 11}, 4623
(1996) [arXiv:gr-qc/9607022].



\bibitem{un}
P.~C.~W.~Davies, S.~A.~Fulling and W.~G.~Unruh, ``Energy Momentum
Tensor Near An Evaporating Black Hole,'' Phys.\ Rev.\ D {\bf 13},
2720 (1976).

\bibitem{bi}
N.~D.~Birrell and P.~C.~W.~Davies, "Quantum Fields in Curved
Space," Canbridge University Press (1984).


\bibitem{massar}
R.~Brout, S.~Massar, R.~Parentani and P.~Spindel, ``A Primer For
Black Hole Quantum Physics,'' Phys.\ Rept.\  {\bf 260}, 329
(1995).


\bibitem{pol}
J.~Polchinski, L.~Susskind and N.~Toumbas, ``Negative energy,
superluminosity and holography,'' Phys.\ Rev.\ D {\bf 60}, 084006
(1999) [arXiv:hep-th/9903228].


\bibitem{lm}
O.~Lunin and S.~D.~Mathur, ``AdS/CFT duality and the black hole
information paradox,'' Nucl.\ Phys.\ B {\bf 623}, 342 (2002)
[arXiv:hep-th/0109154].

\bibitem{lm2}
O.~Lunin and S.~D.~Mathur,
 ``Statistical interpretation of Bekenstein entropy for systems with a
stretched horizon,'' Phys.\ Rev.\ Lett.\  {\bf 88}, 211303 (2002)
[arXiv:hep-th/0202072].



\bibitem{ma3}
S.~D.~Mathur, ``Where are the states of a black hole?,''
arXiv:hep-th/0401115.






\bibitem{hamed}
N.~Arkani-Hamed, H.~C.~Cheng, M.~A.~Luty and S.~Mukohyama, ``Ghost
condensation and a consistent infrared modification of gravity,''
arXiv:hep-th/0312099.




\bibitem{thhol}
G.~'t Hooft, ``Dimensional Reduction In Quantum Gravity,''
arXiv:gr-qc/9310026.

\bibitem{suhol}
L.~Susskind, ``The World as a hologram,'' J.\ Math.\ Phys.\  {\bf
36}, 6377 (1995) [arXiv:hep-th/9409089].




\bibitem{iw}
K.~Iwasawa {\it et al.}, ``The variable iron K emission line in
MCG-6-30-15,'' arXiv:astro-ph/9606103.





\bibitem{bh}
R.~Narayan, ``Evidence for the Black Hole Event Horizon,''
arXiv:astro-ph/0310692.

\bibitem{fs}
W.~Fischler and L.~Susskind, ``Holography and cosmology,''
arXiv:hep-th/9806039.

\bibitem{bf}
T.~Banks and W.~Fischler, ``An holographic cosmology,''
arXiv:hep-th/0111142.

\bibitem{bou}
R.~Bousso, ``Positive vacuum energy and the N-bound,'' JHEP {\bf
0011}, 038 (2000) [arXiv:hep-th/0010252].



\bibitem{banks1}
T.~Banks, ``Breaking SUSY on the horizon,'' arXiv:hep-th/0206117.

\bibitem{banks2}
T.~Banks,
 ``Supersymmetry, the cosmological constant and a theory of quantum  gravity in
our universe,'' Gen.\ Rel.\ Grav.\  {\bf 35}, 2075 (2003)
[arXiv:hep-th/0305206].




\bibitem{sn1}
J.~L.~Tonry {\it et al.}, ``Cosmological Results from High-z
Supernovae,'' Astrophys.\ J.\  {\bf 594}, 1 (2003)
[arXiv:astro-ph/0305008].

\bibitem{sn2}
W.~J.~Percival {\it et al.}, ``The 2dF Galaxy Redshift Survey: The
power spectrum and the matter content of the universe,'' Mon.\
Not.\ Roy.\ Astron.\ Soc.\  {\bf 327}, 1297 (2001)
[arXiv:astro-ph/0105252].

\end{thebibliography}
\end{document}